\documentclass[1opt]{article}
\usepackage{graphicx,amsmath,amsfonts,amssymb,dcolumn,amsthm,euscript}
\def\bea{\begin{eqnarray}}
\def\eea{\end{eqnarray}}
\def\be{\begin{equation}}
\def\ee{\end{equation}}
\usepackage{color}

\newcommand\nn{\nonumber} 
\newcommand{\bq}{\begin{equation}}
\newcommand\eq{\end{equation}}

\def\bar{\overline}
\newcommand\pa{\partial}

\def\d{{\rm d}}

\newcommand\del{\delta}

\usepackage{slashed}

\def\demi{\frac{1}{2}}

\begin{document}
\title{\textbf{ 
 {   Unimodular Gauge    in  Perturbative  Gravity and Supergravity } \\ 
 }}
\author{\textbf{Laurent Baulieu
 }
\thanks{{\tt baulieu@lpthe.jussieu.fr}
}
\\\\
\textit{
%
LPTHE, Sorbonne Universit\'e, CNRS} \\
\textit{{} 4 Place Jussieu, 75005 Paris, France}\\
}

\def\blue{  \color{blue}}

\def\blue{  \color{black}}

\def\red{  \color{red}}
    \def\betar {{\red{\beta}}}
          \def\betar {{{\beta}}}

\date{
$ $
}
\maketitle
\begin{abstract}
  This        paper     explains the Unimodular gauge fixing of gravity and supergravity  in  the framework  of 
    a    perturbative BRST construction.
     The unphysical sector contains   additional BRST-exact quartets  to  suppress possible ambiguities and impose both the 
     Unimodular gauge fixing condition on the metric and a gauge condition for
     the reparametrization   symmetry of the unimodular part of the metric.  
      The  Unimodular gauge choice of the metric  must be completed     by a $\gamma$-Traceless gauge condition for the    Rarita--Schwinger field  in the case  of supergravity. This gives  an interesting new class of gauges for gravity and supergravity. 

\end{abstract}

\def\s{{s_{stoc}}}
\def\hl{{\hat l}}

\def\l{\lambda}
\def\ll{{\hat \lambda}}
\def\th    {\theta}
\def\thh{{\hat \theta}}
\def\ee{{\epsilon}}
\def\eeh{{\hat \epsilon}}
\def\dd{\delta}
\def\ddh {\hat \delta}

\def \pa{\partial}
 \def\l{{\lambda}}
  \def\g{{\gamma}}
   \def\T{{t}}
   \def\t{\tau}
  \def \d{\delta}
 \def\d{{\rm d}}
 \def\x{{q}}

 \def   \hg { {\hat g}}
  \def\g {{\sqrt{g}}}
  
  \def\bx  {
   {  {{\bar \xi}}^{(01)   } } }
   
  \def\bxm {
   {  {{\bar \xi}}^{(01)  \mu} } }
   
    \def\bxn {
   {  {{\bar \xi}}^{(01)  \nu} } }
   
     \def\L  {
   {  {{L }}^{(00)   } } }
   
   \def\Lx{     Lie_{\xi   }  }  
  
      \def\s{  \hat  s   }
          
    \def\hg{    {\hat g}} 
    
       \def\g{    {\sqrt  {g}  }}
 
 \def\v{ {\varphi}}
     \def\s{     s   }
     \def \vx {\vec{x}}
     \def\V{   {{\Phi}    }}
     \def \Vx { \V (\vx)}
     \def\vv{   {\vec {v}    }} 
     \def\vV{   {\vec {V}    }}
  \def \vX {\vec{X}}
  
  \def\Gam{\Gamma}
 \def\ol{\hat}
  \def\Sig{\Sigma} 
  
 \newpage

\section {Introduction}
Albert Einstein    recognized  as early as in 1916 that  there is a preferred gauge  in classical gravity. 
He recommended    the choice  of  a system of coordinates such that the determinant  of the  space-time metric
$g_{\mu\nu}$  is    locally unimodular. This means the  gauge choice $-\det(g_{\mu\nu})=1$ for  solving  the Einstein equations of motion~\cite{Einstein}.   Since this epoch, there has been some interesting activities     about  the concept of the Unimodular gauge.  The word   ``Unimodular gravity" has actually become quite common.  The  non-exhaustive  series of papers 
\cite{Henneaux}\cite {Alvarez1}\cite{Padilla} \cite{Bufalo}\cite{Alvarez2}\cite{Oda1}\cite{Oda2}\cite{Percacci}\cite{Nagy}\cite{Martin}\cite{Vikman}
and  references therein    address   interesting  questions     related to this domain.  
 A priori, the  so-called Unimodular gravity has    a  different  physical content than the standard Einstein theory.  When looking at the literature,  
  there are mainly two formulations: one  that imposes $g\equiv -\det(g_{\mu\nu})=1$ as a gauge choice and another one  that    imposes the constraint $\sqrt g=1 $.  Some confusion  is   spread  around   these formulations, although their difference is actually quite  clear.  
  
Working in the  ``Unimodular gauge"   $ g = 1$ for the     Einstein theory  
is   nothing but a possible choice, maybe unfamiliar  and difficult to   enforce, but formally equivalent to any  other   gauge choice.  The  BRST methodology to enforce  this
gauge choice      is the 
subject of this paper. It  exhibits interesting non trivialities that we find worth being published.
The end of  this introduction  sketches   physical   motivations for using this gauge.
Maybe the  most striking one is that  the gravity observables  can be represented as functionals of the  unimodular part of the metric, 
because of the physical  redundancy between  metrics related by a Weyl transformation.
This last property was  underlined  in a different way in  the classical  theory  in \cite{York}. 

 In contrast, the ``Unimodular gravity" means that one changes  the theory by varying classically  the 
Einstein-Hilbert action by only considering variation of   metrics with $\sqrt g=1$. One motivation of 
the ``Unimodular gravity" is that the cosmological constant is   introduced as a   constant of integration that can be chosen at will,
while  the  ``Unimodular gauge"   fixes the cosmological constant as a parameter of the Lagrangian   from the beginning.
 

 Imposing $\sqrt g=1$  is a   well-defined classical local gauge condition for gravity  made possible by      the reparametrization  invariance  of the  theory.
  Thus, as a matter of principle, there should be no ambiguity to define  a   perturbative quantum field theory   of   gravity in this gauge, at least semi-classically.
       A  solution     must exist  perturbatively for 
   quantizing gravity by imposing  the unimodular   condition  $\sqrt g=1$ on $g_{\mu\nu}$  and    gauge fixing   afterwards
   the residual  reparametrization invariance
   of its unimodular part $\hat g_{\mu\nu}$. If supergravity is involved,   the Unimodular  condition   on the metric also implies a $\gamma$-traceless condition  on the spin $3/2$  Rarita--Schwinger field  $\Psi_\mu$ and  
   the  residual local supersymmetry of its  pure spin $3/2$  part $\hat \Psi_\mu$ must be further gauge fixed.

 This paper
is  thus aimed at building a local quantum Lagrangian  that  defines  gravity  and supergravity unambiguously in the  Unimodular gauge, at least for defining   a  consistent  perturbative BRST invariant  quantum field theory.
A BRST exact   gauge fixing  action     will  be build         that  enforces  consistently the Unimodular gauge condition, to be    
   added  to  the  Einstein action (and the Rarita--Schwinger action).  
    We don't   fear a possible anomaly  for this process in the $d=4$ case,  because  consistent $4$d  gravitational anomalies cannot possibly exist due  to  the structure of the $SO(3,1)$ Lie algebra. 
   
 For any given choice of a classical gauge function,  getting  a BRST symmetry  invariant  gauge fixing   is  necessary to possibly enforce all relevant Ward identities that   define the quantum theory eventually.   Gravity    is non renormalizable by power counting but, presumably,    the Unimodular gauge fixing procedure can be made   stable under radiative corrections  by introducing the   needed counterterms that are compatible with the Ward identities in this gauge,   order by  order in perturbation theory. 

  The Unimodular gauge  quantum Lagrangian   built in this work  makes   explicit  some particularities  of  diffeomorphisms with a divergence-less vector parameter.  One of its  subtleties is that    a formal    Faddeev--Popov  gauge fixing of     Unimodular  metrics     provides a singular determinant with  a  ghost  of ghost phenomenon.  
To control this phenomenon,   techniques analogous  as those used   to currently define  TQFT's with a gauge invariance are needed.   
This provides  a localisation of   fields  around  their unimodular components  with  a remaining degeneracy     to be further   fixed in a BRST invariant way. It follows an enlargement of  the standard   BRST  field content  of perturbative gravity  with additional  BRST trivial quartets 
to define  the Unimodular gauge.{\blue The  Unimodular gauge  fixing   in first formalism and the  expression of the spin connection in this gauge  will be also  discussed.}
   
Although  this work is self contained, it has  a  hidden   motivation   that is the stochastic quantization of gravity. The latter    remedies    the absence  of a well-defined    Lorentz time evolution    in quantum gravity  by the  stochastic  quantization time as the variable that  orders
 the    non-perturbative  quantum gravity phenomena. In this framework, \cite{bcw}    indicates that the conformal factor of the metric behaves as  a spectator,  while  the relevant  non  trivial aspects of the quantum gravity dynamics are carried by the unimodular part of the metric. But \cite{bcw}~also predicts that,   at  the perturbative level,    the   limit at infinite stochastic time  of 
 the stochastically  quantized   gravity  is   the well-defined (modulo UV questions) standard $4$d perturbative quantum theory, for which the Lorentz time can be defined. To show this  result,  \cite{bcw}  uses  the decomposition of the metric in its unimodular component and conformal factor.  
 It    thus   appears necessary  to dispose of a  precise   construction of   semi-classical   gravity in  the unimodular gauge,   the subject    
of this paper. 

 Interestingly,   having  a well-defined     
 perturbative quantization of gravity   in the  unimodular gauge  makes contact  with the work of York~\cite{York}, who showed    that what the classical Einstein equations  truly propagate   are the  equivalence classes of metrics defined modulo Weyl transformations.
Solving the  Einstein equations   is a Cauchy problem. York pointed out that,  taking as initial conditions two metrics related by a Weyl transformation,  their  evolution at any given future time 
  provides  two metrics that are also related by a Weyl transformation. This fact holds true  although the gravity equations of motions are not Weyl invariant~\cite{York}. This  makes the principle of gauge invariance  and the definition of   observables more subtle in gravity than in Yang-Mills  and $p$-form 
gauge invariant theories.
  Showing that one can  gauge fix the  metric to be unimodular in  a BRST invariant way   is  a way to generalize   at the quantum level    the   classical arguments of York, since   the set   of the Weyl  classes of metrics can be represented by the set of unimodular metrics.  

Our work     suggests that the (super)gravity observables should  be defined as the functionals of unimodular metrics (and gravitino $\gamma$-Traceless components)  although the (super)gravity  action  is  not Weyl invariant.  This property is very natural when one works in the Unimodular gauge. Since the BRST invariance  ensures that
the same physics can be computed  with any other (well-defined) choice of gauge, the same conclusion  must be true in other gauges. The expression of observables may then 
occur with more complicated  expressions.

 \section{Pure Gravity}
 
 \subsection {Improved BRST symmetry for the Unimodular gauge}

The current  method  to perturbatively  gauge fix gravity in   Lagrangian formalism is by  introducing  a BRST symmetry operation $s$   acting on   the metric field $g_{\mu\nu}(x)$  and the vector   ghost field  $\xi^\mu(x)$ of the reparametrization symmetry. The  covariance  of 
the     BRST trivial pair   made of  a reparametrization   antighost   and  a Lagrange multiplier depends on the gauge condition one wishes to use.  Choosing 
the    gauge function $\pa_\nu g^{\mu\nu}$,   the  anticommuting antighost  and  commuting Lagrange multiplier are both vector fields  $\bar \xi^\mu(x)$ and  $b^\mu(x)$.
The  BRST symmetry   is         defined by the following    graded  differential operator  $s$  acting on  the  gravity BRST  multiplet fields
     \bea \label{brstg}
      sg_{\mu\nu}&=& \Lx      g_{\mu\nu}  \nn \\
     s \xi^\mu &=&\xi^\nu   \pa_\nu   \xi^\mu\nn\\
      s\bar \xi^\mu &=&b^\mu \rm \nn\\
       s b^\mu&=&0.
     \eea  
   One  has   $[s,\pa_\mu]=0 $ and  the nilpotency   $s^2=0$.  t'Hooft and Veltman      defined     the perturbation expansion of quantum gravity  in the   de-Donder gauge   by adding     the  $s$-exact  term $s( \bar \xi^\mu     \pa^\nu g_{\mu\nu})$ to  the Einstein  action~\cite{VH}
 \bea
 \label{ddg}
  L_{Einstein} \to   L_{Einstein} + s( \bar \xi^\mu     \pa^\nu g_{\mu\nu})
  =  L_{Einstein} +b^\mu \pa^\nu g_{\mu\nu} - \bar \xi^   \mu   Lie_   \xi  \pa^\nu g_{\mu\nu} .
\eea
  They used    the  Feynman rules for the  metric  and  the ghosts and antighosts that   stem from  the local  action~(\ref{ddg}).  Their gravity Ward identities are implied by    the symmetry~(\ref{brstg}) where $  {\cal L}_\xi   g_{\mu\nu}=   g _{\rho\mu}   \pa_\nu  \xi^\rho
+  g _{\rho\nu}   \pa_\mu  \xi^\rho   + \xi^\rho \pa_\rho \hat g_{\mu\nu}$.
  
    Using an Unimodular    gauge choice with $ \g\equiv 
\sqrt{   - \det    g_{\mu\nu}  }=1$ seems impossible  with  only the standard  Fadeev--Popov fields:   the $4$     conditions   $\pa_\mu g^{\mu\nu}=0$   exhaust  the   possibilities   allowed by the   Lagrange  multiplier~$b^\mu$.

 In fact, 
something more refined than the standard Faddeev--Popov construction must  be done to define the    gauge fixing 
of  $  g_{\mu\nu}$ to   its  unimodular part  $\hat g_{\mu\nu}$, defined as  
 (here $d=4$)\bea \hat g_{\mu\nu}\equiv  g_{\mu\nu}/   {\sqrt{ g} }^{\frac {2}{d}}, \eea
with  a further  gauge fixing of the    reparametrization symmetry of $\hat g_{\mu\nu}$,{  which    satisfies   \bea\label{diffhat}
 s\hat g_{\mu\nu}\equiv{\cal L}_\xi \hat g_{\mu\nu}= \hat g _{\rho\mu}   \pa_\nu  \xi^\rho
+\hat g _{\rho\nu}   \pa_\mu  \xi^\rho +\xi^\rho \pa_\rho \hat g_{\mu\nu}-
\frac 2 d \hat g_{\mu\nu}   \hat \nabla_\rho \xi^\rho.
\eea
 The value of the  coefficient of the  last term    in  (\ref{diffhat})  ensures that   $
  \hat g^{\mu\nu} s\hat g_{\mu\nu} =0$ consistently with $\det  \hat g_{\mu\nu} =1$.}
 
The clarification    comes by considering $\hat g_{\mu\nu}$ and $g$  as the independent quantum  field    variables,  with reparametrization transformations defined   by  (\ref{diffhat}) for  $\hat g_{\mu\nu}$  and   $s \sqrt g  = \nabla _\mu\xi^\mu$.         This  generalizes 
   for $d>2$
the decomposition  of a  $2$d metric in its  Beltrami parameter  and its conformal factor~\cite{bb}.
The   off-shell  decomposition  of    $  g_{\mu\nu}$ in $\hat g_{\mu\nu}$ and $\sqrt g$   is     justified because the  variations 
  of $  g_{\mu\nu}$ are not   irreducible Lorentz tensors and      split   into   trace and traceless components. 
  In fact,   an off-shell decomposition  of   any given Lorentz tensor fields in irreducible representations  should   be done    systematically   for      spin values larger than~$1$.  For spin~$3/2$,   the Rarita--Schwinger field  $ \Psi_\mu$  must be split  
in its   $\gamma$-Trace       and $\gamma$-Traceless irreducible components, and so-on.

The 
 gauge fixing problem   of  gravity   in  the Unimodular gauge   draws    us quickly deeper     in the
BRST symmetry formalism than 
 the  Yang--Mills theory and, more generally, than  the theory of    $p$-form fields  whose field  variations  belong to
irreducible Lorentz 
 representations.  {  The reason is that   if one  formally applies the Faddeev--Popov  method and   impose  both gauge conditions    $\pa_\nu \hat g^{\mu\nu}=0$ and  $\sqrt g=1$, (which  make   sense  classically), the situation becomes confusing. 
   The   Unimodular gauge condition    $\sqrt g=1$   equates  the Einstein action   density $\g R(g_{\mu\nu})$ as  $R|_{g_{\mu\nu} =\hat g_{\mu\nu}}$ and the later term is   invariant under  all restricted diffeomorphisms   with a  divergent-less   vector field   parameter $\xi^\mu$ with  $\nabla_\mu\xi^\mu=0$  according to~(\ref{diffhat}). 
   But for such a  vector field,       one has 
 \bea\label{co}
 s \sqrt g \Big |_{\sqrt g=1}= \pa_\mu\xi^\mu=0.
 \eea 
This justifies the necessity   of  separating the "longitudinal" components  of $\xi^\mu$ (satisfying $\pa_\mu  \xi^\mu =0 $) from its "transverse" component.
 The same  must be done for the    antighost  partner $\bar \xi^\mu$  of $\xi^\mu$ to understand the further gauge fixing of  $R|_{g_{\mu\nu} =\hat g_{\mu\nu}}$.     If it can be done, $\hat g_{\mu\nu} $ and $g$ can be truly treated as  independent fields, with a well-defined       path integral  measure  in a consistent BRST approach for the Einstein  action  in the   Unimodular~gauge.   
   }

The use of  
ghost  and/or   antighost  fields  defined   modulo some degeneracy is often   done by introducing      ghosts of ghosts. In our case, the use of ghosts of ghosts   
will correct  very concretely   the wrong statement  that       the~5~conditions  $\pa_\nu \hat g^{\mu\nu}=0,\sqrt g=1$  might  imply    an over-gauge fixing. The longitudinal and transverse  components  of  the  auxilary field $b^\mu=s \bar\xi^\mu$  
must be also separated as those of $\xi^\mu$ and $\bar\xi ^\mu$.   The  longitudinal component of~$b^\mu$ ~may need    a BRST invariant  gauge fixing.  
The current  understanding 
 of topological quantum field theories with gauge symmetries     involving systematically ghosts of ghosts       can be used as a road map. It justifies the introduction   of     extended   BRST symmetries involving  new fields   organized under the form of BRST-exact quartets.
  Such quartets  count altogether for zero degrees of freedom  and  solve in general all   issues about ghosts with an internal degeneracy.   Their field components often 
play the role of Lagrange multipliers.    For   gauge fixing   the unimodular part of the metric such quartets will allow   the construction of    a BRST invariant  path  integral  with a  functional measure 
 using   $\hat g_{\mu\nu} $ and $\g$ as    fundamental   fields. 
 
 One   thus completes  the ordinary  BRST system  in Eq.~(\ref{brstg})
 by addition of  the trivial BRST quartet
    \bea    \L,  \ \eta  ^ {(10)},\bar \eta  ^ {(01)},\ b  ^ {(11)}   .
 \eea
    The    scalar  bosonic  fields  $L,b$ and       fermionic   fields  $\eta,\bar \eta$   count altogether for zero=1+1-1-1  degrees of freedoms in unitary relations provided  their dynamics is governed  by  an s-exact  action defining invertible propagators.
    
        Having  available  this  extra set of   unphysical fields     is  exactly what one  needs to get   a Lagrangian with    invertible propagators    in     the Unimodular gauge, with  a BRST invariant gauge fixing of   zero modes that otherwise  would spoil the definition of gravity by a path integral   in the  Unimodular gauge.
        Eventually,  a path integral with a measure     depending  only on  the unimodular part of the metric will be obtained. One can interpret this result as the quantum  generalization of the classical  prescription  of Einstein~\cite{Einstein}.

       The following diagram displays suggestively all    necessary ghosts, antighosts and Lagrange multipliers\footnote{ The notation   $\phi^{g,g'}$  means that the field $\phi^{g,g'}$ carries ghost number  $g$ and antighost number $g'$ for a total {\it net ghost number } $G=g-g'$.  $\phi^{g,g'}$ is a boson if    $G$ is even   and a fermion if $G$ is odd. We often skip these ghost and antighost indices in the formula.}
        \def\red{\color{red}}
\bea\label{spg}  \quad \quad
\begin{matrix}
     &     g_{\mu\nu }=(\hg_{\mu\nu},  \g),   \L &   & &   \\
 \swarrow  \  \ \ \  \swarrow& &    &  &   & & \\
    \xi^{\mu(10)},   \eta  ^ {(10)}  &    &    \bar  \xi_\mu  ^ {(01)},   \bar  \eta  ^ {(01)} & \\
       &   \swarrow \ \ \  \swarrow &  &    \\ 
     &     b_\mu  ^ {(11)},   b  ^ {(11)}  &    &    \\ 
     &   &  &    & &   \\
  1   &   0 &     -1  & 
\end{matrix}.  \eea
The  numbers  $-1,0,1$   in  the  bottom line    indicate  the   net ghost number of      fields that are   aligned  vertically   above  each number.
The     BRST transformations  that generalizes     (\ref{brstg}) are
 \bea      \label{aux} 
  sg_{\mu\nu}&=& \Lx  \  g_{\mu\nu} \nn\\
   s\xi^\mu&=&   \Lx  \xi^\mu 
   \nn \\  
\s\bar \xi    ^\mu   &=& b     ^\mu    \quad  \quad \quad\quad \quad\quad \quad
    \s b  ^\mu   =0
     \nn\\
   \s L   &=& \eta      \quad \quad \quad \quad \quad\quad \quad
   \  \s \eta      =0
     \nn\\
      \s\bar\eta      &=& b    \quad \quad \quad\quad \quad\quad \quad
  \ \   \s b   =  0  .
      \eea
    One    still  has $[s,\partial _\mu]\equiv 0$, $\{ s,\d\} \equiv 0$ and     
$  { s^2  =0 }$ on this extended set of fields.
 In fact  $d$, $s$,  $i_\xi$, $Lie_\xi=[i_\xi,d]$
 build    a system of nilpotent graded differentials  operators.\footnote{ 
$\hat s=s-L_\xi$  is nilpotent as $s$ in the absence of local supersymmetry because in this case  $  s\xi=   \Lx \  \xi$.
In supergravity, ${\hat s}^2= i _\Phi\neq 0$, where $\Phi^\mu  =\chi\gamma^\mu\chi $ is  the vector field quadratic in the commuting  supersymmetry ghost $\chi$ \cite{bb}.} 
The last two lines in Eqs.~(\ref{aux})  identify    $L,\eta,\bar\eta,b$ as the elements  of    a BRST exact quartet. The commuting scalar $b $ 
is   an additional    scalar  Lagrange multiplier with ghost number 0.  Both  anticommuting scalar     $\eta$,   $\bar\eta$    
are  odd Lagrange multipliers with   ghost numbers $-1$ and $1$.{

  \subsection{The BRST invariant quantum Einstein Lagrangian for the Unimodular gauge}
  Define now  a class of  BRST invariant gauge fixing actions with the  gauge functions 
$\pa_\rho    { \hg}^{\rho \nu} $ and $\sqrt g-1$. 

 Using the definition of $s$ in  (\ref{aux}), one   can  complete  $ \int dx   \g R(g_{\mu\nu})  $ by  addition of an $s$-exact term. One defines
    \bea\label{Egf}
\int dx {\cal L}^{\rm BRST\ inv.}_{\rm gauge  \ fixed}=  \int dx \Big(   \g R 
 +s \     \Big[\     \bar \xi^{ \mu} ( \hat g _{\mu\nu}\pa_\rho    { \hg}^{\rho \nu}  +\gamma \pa_\mu  L+\frac{\alpha}{2}b^\mu) 
 +\bar \eta ( \g-1) 
 \Big]
\Big).
 \eea
 The range of these Unimodular  gauges is parametrized     by  all possible choices for the  gauge parameters $\gamma\neq 0$ and  $\alpha$. 
 Observables are  the elements of    the cohomology of $s$.  Their expectation values  are    independent on the choice of $\alpha$
 and   $\gamma\neq0$. 
We will consider the case    $\alpha=0$ and   $\gamma=1 $.   
 Expanding the $s$-exact term  yields
  \bea     \label{Egff}
\int dx {\cal L}^{\rm BRST\ inv.}_{\rm gauge \ fixed}
(  g_{\mu\nu}, L, b_\mu,\xi^\mu,
\bar\xi^\mu,
\eta,\bar\eta,b)
=  \int dx  \Big (   \g R (g_{\mu\nu})+
     b ^\mu (  \hat g _{\mu\nu}\pa_\rho    { \hg}^{\rho \nu}
      +\pa_\mu  L ) 
 +b  ( \g-1)
 \nn \\
 -
\bar \xi  ^\mu  (   \hat g_{\mu\nu}   \pa_\rho    {\rm {Lie}} _\xi   { \hg}^{\rho\nu}   + 
({\rm {Lie}}_\xi \hat g _{\mu\nu})\pa_\rho    { \hg}^{\rho \nu} 
-\bar \xi  ^\mu \pa_\mu  \eta 
 +\bar \eta     \nabla _\nu \xi^\nu \Big ).
 \eea
 ($ \rm{ Lie }_\xi \hat g_{\mu\nu}$ is expressed in~(\ref{diffhat}).) The  equation of motion  of the auxiliary field $b$  imposes  $\sqrt g=1$,  ie  $g_{\mu\nu}=\hat g_{\mu\nu}$,~everywhere in (\ref{Egff}). Its elimination yields  the   BRST  invariant action    of gravity in the Unimodular gauge  
\bea\nn\label{LBRSTg}
 I^{\rm BRST}_{\rm Unimodular  }\Big[   \hat g_{\mu\nu}, L, b_\mu,\xi^\mu,
\bar\xi^\mu,
\eta,\bar\eta\Big]=   \qquad\qquad\qquad\qquad \qquad
\\
 \int dx  \Big (  R (\hat g_{\mu\nu})+
     b ^\mu (\   \hat g _{\mu\nu}\pa_\rho    { \hg}^{\rho \nu}
      +\pa_\mu  L \ ) 
  -
\bar \xi  ^\mu  ( \   \hat g_{\mu\nu}   \pa_\rho    {\rm {Lie}}_\xi   { \hg}^{\rho\nu}   + ({\rm {Lie}}_\xi \hat g _{\mu\nu})\pa_\rho    { \hg}^{\rho \nu} \ )
-\bar \xi  ^\mu \pa_\mu  \eta 
 +\bar \eta     \nabla _\nu \xi^\nu  \Big).
 \eea
{\blue{The BRST invariant gauge-fixed  action       (\ref{LBRSTg}) and  the associated         non linear coupled differential equations  of motion  may  look  complicated at first sight.  It  is worth   explaining the role of all terms in $I^{\rm BRST}_{\rm Unimodular  }$.
 $R ( {  \hat g_{\mu\nu})}$~stands for  $R ( {    g_{\mu\nu})}$ where one replaces  
  $   g_{\mu\nu}$ by $ \hat g_{\mu\nu}$.
   A reparametrization invariance   remains for $R ( {  \hat g_{\mu\nu})}$,  but with  the constraint $\nabla_\mu \xi^\mu=0$ on    the longitudinal part of the vector $\xi^\mu$,  as implied by the transformation  law  ~(\ref{diffhat}). The reading of  $I^{\rm BRST}_{\rm Unimodular  }$ shows that this constraint   is the fermionic equation of motion of $\bar\eta$.
   Before the   gauge fixing of $\sqrt g=1$ by  the equation of motion  of      $b=s\bar \eta$,   $\hat g_{\mu\nu}\equiv   g_{\mu\nu}/{g}^{\frac 1 d}$ is a composite  of $g_{\mu\nu}$ and $\sqrt g$. After the   gauge fixing $\sqrt g=1$, $\hat g_{\mu\nu}$ becomes an  independent field     with~$\frac {d(d+1)}{ 2}-1 $  degrees of freedom whose covariance is defined by  (\ref{diffhat}). This makes consistent 
the   approach        that     identifies     the   $\frac {d(d+1)}{ 2}-1 $ independent degrees of freedom  of the  unimodular matrix    
$\hat g_{\mu\nu}$   
 and the single one carried   by  $  \sqrt g $ (or equivalently   by $\phi$,      $g \equiv \exp  {  -} 2 \phi $) as the~$\frac {d(d+1)}{ 2}$~independent  fundamental fields of gravity. This  proposition  makes sense classically and our  BRST construction      verify that    it remains true  at the  quantum level. 
 The (lesser relevant)  gauge fixing of  $\hat g_{\mu\nu}$~involves the  auxilary field    $b^\mu$ and   the  propagating field~$L$. $b^\mu$  becomes   a Lagrange multiplier when   $\alpha=0$.    
  Its    equation of motion  enforces  in a BRST invariant way  the  $\phi$ independent condition on~$\hat  g _{\mu\nu}$ 
\bea
\pa_\rho    { \hg}^{\rho \mu}
      + \hat g ^{\mu\nu} \pa_\nu  L =0
\eea
 The  path integration  over all possibilities over the  field $L$     
   avoids an over gauge fixing if one performs    the path integral of $\exp -I^{\rm BRST}_{\rm Unimodular  }$ over all metrics.
The   BRST invariant   action~(\ref{Egf})~is a  quadratic form  of the anticommuting fields $\xi^\mu$,  $\bar \xi^\mu$, $\eta$ and~$\bar\eta$. These ghosts  sandwich     local operators. 
Their   functional integral from  $\exp -I^{\rm BRST}_{\rm Unimodular  }$ determines a now well-defined product of Faddeev--Popov determinants  associated to both gauge functions  $\sqrt g -1 $ and $\pa ^\mu  \hat g_{\mu\nu}$.  
 If one    reinstall the  general   $\gamma\neq1$   gauge parameter dependence  of (\ref{Egf}), the free ghost   field     quadratic approximation of the action  is
\bea \int dx \Big (-g_{\mu\nu}  \bar \xi^\mu\pa^2 \xi^\nu  + (\bar \eta  +\frac{2- d}{ d }  \pa \bar  \xi ) \pa    \xi +\gamma \eta \pa \bar \xi\Big). \eea
Its    invertibility and the absence of   zero modes for the ghosts  imply  $\gamma\neq 0$.  This   justifies our   choice $\gamma=1$ and  the    use   of all the elements  of  the   quartet $(L,\eta,\bar \eta,b)$ for  defining consistently the unimodular gauge.

All this   establishes  that         $I^{\rm BRST}_{\rm Unimodular  }$ is truly the  local 
  BRST invariant   action that   enforces at the quantum level the classically admissible  gauge   functions~$\sqrt g$~and $\pa_\rho    { \hg}^{\rho \nu}$ for   gauge fixing the  reparametrization  invariance of the Einstein action.
 It is instructive enough to observe that the variation   with respect to $\hat g_{\mu\nu}$  of    $\int dx  R (\hat g_{\mu\nu})$~gives the following contribution to the  
 $\hat g_{\mu\nu}$  equation of motion  of the complete BRST invariant action \bea\label{tr}
 \frac  {\delta }{\delta   \hat  g_{\mu\nu}    }    \int dx  R (\hat g_{\alpha\beta})
  =   R_{\mu\nu}   ( \hat  g_{\alpha\beta})     -\frac1 2\ol   g_{\mu\nu}  R(  \hat  g_{\alpha\beta}). 
   \eea
 One recognizes in the right hand side the Einstein tensor
 $E_{\mu\nu}  \equiv R_{\mu\nu}       -\frac1 2    g_{\mu\nu}  R  $
 with  $g_{\mu\nu}$     replaced by its unimodular component $\hat g_{\mu\nu}$, with the already mentioned reparametrization invariance. 
 In fact, following     \cite{Thomas199},  \cite{bcw}  indicates     how to      
   express   all gravity  tensors in term of   $\hat g_{\mu\nu}$ and $\phi$, using 
   the important      Christoffel symbols    decomposition
\bea \Gam^\mu_{\nu\rho}=\ol\Gam^\mu_{\nu\rho}+\varSigma^\mu_{\nu\rho}, \quad \rm{where}\quad 
\ol\Gam^\mu_{\nu\rho}\equiv\demi\ol g^{\mu\alpha}(\pa_\nu\ol g_{\alpha\rho}
+\pa_\rho \ol g_{\alpha\nu}-\pa_\alpha \ol g_{\nu\rho}), \quad 
\varSigma^\mu_{\nu\rho}\equiv \delta^\mu_\rho\pa_\nu\phi+
\del^\mu_\nu\pa_\rho\phi-\ol g^{\mu\alpha}\ol g_{\nu\rho}\pa_\alpha\phi. 
\eea
 The ``hat  covariant derivative'' $\hat\nabla_\mu$ is
 $\nabla_\mu$ where  the Christoffel   $\Gam^\mu_{\nu\rho} (  g_{\alpha\beta})$
 is replaced by~$\hat\Gam^\mu_{\nu\rho}(\hat g_{\alpha\beta})$.
$\hat R_{\mu\nu}$~and $\hat R$ are defined  as  $R_{\mu\nu}$
and $R$, but using the hatted quantities $\hat\Gam^\mu_{\nu\rho}$ and
$\ol g_{\mu\nu}$ (for instance   $\hat R \equiv R ( {  \hat g_{\mu\nu})} $).   \cite{bcw} shows
\bea
R_{\mu\nu}(g_{\alpha\beta})&=&\hat R_{\mu\nu}(\hat g _{\alpha\beta})-(d-2)\hat\nabla_\mu \pa_\nu \phi-\hat g_{\mu\nu}
\hat\nabla_\alpha \hat g^{\alpha\beta}\pa_\beta\phi+(d-2)\pa_\mu \phi\pa_\nu
\phi-(d-2)\hat g_{\mu\nu}\pa_\alpha\phi\hat g^{\alpha\beta}\pa_\beta\phi \nn \\
R(g_{\alpha\beta})&=&g_{\mu\nu}R^{\mu\nu}(g_{\alpha\beta})=\exp(-2\phi) \Big(\ol R (\hat g_{\alpha\beta})-2(d-1)\ol g^{\mu\nu}\big(
\ol\nabla_\mu\pa_\nu\phi+\frac{d-2}{2}\pa _\mu\phi\pa_\nu\phi\big)\Big).
\eea
 \cite{bcw}  also computes the   traceless component  of the   variation with respect to $ \hat g_{\mu\nu}$  of   $\int dx R(\hat g_{\mu\nu})$
 \bea\label{scale}
E^T_{\mu\nu}&=&\ol R_{\mu\nu}-\frac1d\ol g_{\mu\nu}\ol R
-(d-2)(\ol\nabla_\mu\pa_\nu \phi -\pa_\mu \phi\pa_\nu\phi)
+\frac{d-2}{d}\hat g_{\mu\nu}\hat g^{\alpha\beta}
\big(\hat\nabla_\alpha\pa_\beta\phi-\pa_\alpha\phi\pa_\beta\phi\big)  \nn\\
&=&\ol{E}^T_{\mu\nu}-(d-2)(\ol\nabla_\mu\pa_\nu\phi-\pa_\mu\phi\pa_\nu\phi)^T.
\eea
 This   decomposition  helps understanding  the meaning   of   Eq.(\ref{tr}) by taking $\phi=0$.
 Eventually,  although~(\ref{LBRSTg}) and its equations of motion may look impressive, everything   relies on   a    consistent   and meaningful construction.}}
 
    Perturbatively,  the consistency of the gauge fixing  provides a  matricial system of invertible propagators for all fields. The unimodular components 
  $\hat g_
 {\mu\nu}$ of  $ g_
 {\mu\nu}$ circulate  in Feynman  diagrams  loops while  $\g$   remains a  spectator field with  some compensations  due to  bosonic   loops  of $L$   and  fermionic loops of  $\eta$ and~ $\bar \eta$.
  All     propagators between the   bosons
  $   \hat g_{\mu\nu},b_\mu, b, L$   and  the  fermions 
$\xi^\mu,\bar\xi^\mu,\eta,\bar\eta$ are     invertible (provided $\gamma\neq 0$)\footnote {
 For  $\alpha\neq 0$, one has a Feynman type propagator for   $\hat g_{\mu\nu} $ and a Klein--Gordon propagator for $L$ after the algebraic elimination of $b^\mu$. There are   mixed propagators between $\hat g_{\mu\nu} $ $b^\mu$ and $L$ as a consequence of the the choice $\alpha=0$.}.
 Thus, the     action~(\ref{Egf}) is well suited   for    a   quantum  description of gravity with the   Unimodular gauge choice $\sqrt g=1$,
  giving a 
 concrete sense to the 
visionary classical prescription   of  Einstein~\cite{Einstein}  as a genuine gauge fixing prescription,  valid also at the quantum level. For     perturbations   around   non-trivial  classical backgrounds, the latter classical fields must be expressed in the Unimodular~gauge.
%
   \subsection{Gravity observables}
   Mean values of observables are  defined as 
  \bea\label{observable}
 < {\cal {O} } (\hat g_{\mu\nu})>
 \equiv
  \int [ d\hat g_{\mu\nu}) ][d\xi^\mu]  [d\bar \xi^\mu]  [dL]
  [d\eta]  [d\bar\eta]
   {\cal {O} } (\hat g_{\mu\nu})    \nn \\
   exp  -\frac{1}{\hbar} 
I^{\rm BRST}_{\rm Unimodular  }\Big[   \hat g_{\mu\nu}, L, b_\mu,\xi^\mu,
\bar\xi^\mu,
\eta,\bar\eta\Big].
 \eea
 If      matter    is coupled,  the gauge fixing    $g_{\mu\nu}=\hat g_{\mu\nu}$ also affects  its energy momentum tensor, 
 which then depends  on $g_{\mu\nu}$ only  
 through $\hat g_{\mu\nu}$.
   Because gravitational anomalies  cannot exits in $d=4$,   the Unimodular gauge can be enforced  order by order at any finite order of perturbation theory, modulo the  necessity of adding  more and more  relevant local  counterterms. The Ward identities should guarantee   the stability of the   gauge~ $\sqrt g=1$.  \def\hPsi {{\hat \Psi}}
    \def\gPsi {{\hat   \slashed\Psi}}
       \def\k {{\chi}}
          \def\kb {{\bar \chi}}
              \def\l {{  \lambda}}
                 \def\lb {{\bar \lambda}}
   \def\a {{a}}
          \def\ab {{\bar \a}}
           \def \spa{        \gamma\cdot \pa}
    \section{Unimodular supergravity}
   We consider the    supergravity $N=1,  d=4$ as an example, but the method is general. We use    the new minimal system of auxiliary fields (a 1-form~$A$ and a 2-form   $B_2$) of  Sohnius and West~\cite{SW} in  the notations of  \cite{BG}. Auxiliary fields are   often necessary for  the nilpotency of the BRST symmetry operator  in supergravity, but  their role  is secondary in this paper\footnote{To  generalize and precisely  incorporate the auxiliary fields dependence  in    the  Unimodular  gauge fixing   of supergravity,  one  can  consistently  use  \cite{BG}, although it is devoted to the different  subject of $N=1,  d=4$ supergravity 
    superHiggs mechanism.}.
 We use  a Lorentz signature. The  flat metric $\eta_{\mu\nu} $ has signature  (-, +, +, +).  The Dirac matrices $\gamma^\mu$ are real and 
$ \gamma^5 \equiv  \gamma^0\gamma^1\gamma^2\gamma^3=\frac {1}{4!}
    \epsilon_ {\mu\nu\rho\sigma}  \gamma^\mu\gamma^\nu\gamma^\rho\gamma^\sigma$. One  
    has $({\gamma^5})^2=-1$,   $\gamma^{5\dagger } =\gamma^{5  }$ and 
   $\gamma^{\mu \dagger } =  \gamma^{0  } \gamma^\mu  \gamma^{0  }$. The Dirac conjugate of a spinor $X$  is    
$ {^*} X \equiv  X^\dagger \gamma^0  $. One   chooses the charge conjugation matrix $C$ to be $\gamma^0$ with  $X^C\equiv   (C  ^*X{} )^T$.  Majorana spinors  have    $4$  real
components  since by definition    $X^C= X^{ }$.
           The  Rarita--Schwinger Lagrangian of  the spin $3/2$ Majorana gravitino  $\Psi_\mu$ is 
  \bea\label{LRS}
  L_{RS}
  =\demi  i
  \epsilon^{\mu\nu\rho\sigma}\ 
   \Psi ^*    _\mu  \gamma^5  \gamma_\nu  D_\rho \Psi_\sigma .
   \eea
     We refer to \cite{bb} as well 
 as to \cite{BG} for  properties  of the covariant derivative $D_\mu =\pa_\mu +\omega_\mu  +A_\mu$ in the new minimal  formulation  of $N=1, \ d=4 $ supergravity  where $\omega$  is the spin connection.  As we already said,  the  dependance in the auxiliary fields $A$ and $B_2$ can be omitted without loss of generality in our discussion.   
 
 The use   of the following       $3/2$   spin projection  operators (as in \cite{PVN} and  \cite{BG}) that    satisfy  all relevant orthogonality  conditions
     make more transparent    the  gravitino gauge fixing.
 \bea
 P^{\frac{3}{2} }     _{ \mu\nu}&=&    \theta_{\mu\nu}
 -\frac 13 \hat \gamma_\mu\hat\gamma_\nu
 \nn\\
 ( P^{\frac{1}{2} } _{11} )    _{ \mu\nu}&=&    
 \frac 13 \hat \gamma_\mu\hat\gamma_\nu \ \ \ \ \ \ 
 ( P^{\frac{1}{2} } _{12} )    _{ \mu\nu}
=\frac {1}{\sqrt 3} \hat \gamma_\mu\Omega_\nu 
\nn
\\
( P^{\frac{1}{2} } _{21} )    _{ \mu\nu}
&=&
\frac {1}{\sqrt 3} \hat \gamma_\nu\Omega_\mu \ \ \ \ \ \ 
 ( P^{\frac{2}{2} } _{12} )    _{ \mu\nu}
=\frac {1}{  3}  \Omega_\mu\Omega_\nu 
\nn
   \\
  \Omega_\mu \equiv\frac {\pa_\mu}{ { \gamma\cdot \pa}}
\ \ \ \ \   \ \ \ \ \ \ \ \   \ \ \ &&
   {\hat{ \gamma}}^\mu\equiv \gamma^\mu- \Omega^\mu
\ \ \ \ \   \ \ \ \ \ \ \ \   \ \ \
   \theta_{\mu\nu}\equiv \eta_{\mu\nu}   - \Omega_\mu  \Omega_\nu.
   \eea
   The free part of  the Rarita--Schwinger Lagrangian,  invariant under  the transformation 
  $\Psi_\mu \to \Psi_\mu + \pa_\mu \epsilon$,  
is 
    \bea  
  L_{RS}^{free}
  =\demi  i
  \epsilon^{\mu\nu\rho\sigma}\ 
  {^*}\Psi_\mu  \gamma^5  \gamma_\nu  \pa_\rho \Psi_\sigma 
  \equiv  {^*}\Psi_\mu   (P^\frac32- P^\frac12_{11})^{\mu\nu} \Psi_\nu  .
   \eea
  An  interesting observation  is that 
 $ L_{RS}^{free}
  =
    \Psi ^*    _\mu  
   (g_{\mu\nu} -  \frac {\pa_\mu\pa_\nu }{\pa^2})  
 \slashed \pa  
   \Psi_\nu  +
  ( ....) \slashed \Psi $. 
  
  Consider   now    the following   algebraic constraint  on  $  \Psi_\mu $  
  \bea\label{gt}
  \slashed \Psi\equiv
  \gamma^\mu \Psi_\mu=0.\eea  
   It can  be  enforced  by  adding  the term   $^*\bar a  \slashed\Psi $ to $ L_{RS}^{free}$ where $\bar a$  is  a fermionic spin~ $1/2$   Lagrange multiplier
  \footnote{Consider the addition of   a mass  term $ im     \Psi ^*    _\mu    \sigma^ {\mu\nu }     \Psi _\nu$ 
 to       
  the free   Rarita--Schwinger   Lagrangian  $L_{RS}^{free}$. One has   $  im     \Psi ^*    _\mu    \sigma^ {\mu\nu }     \Psi _\nu
    =   m   \Psi ^*    _\mu  \Big (
       P^\frac32- P^\frac12_{11} 
       -\sqrt 3(P^\frac12_{11} +P^\frac21_{11} 
     \Big)
   \Psi_\nu\nn$ .
Then,  $L_{RS}^{free, m }$ in (\ref{gfrs}) generalizes into  
 $
  L_{RS}^{free, m }
  =
   \hat\Psi ^*  _\mu  
   \ (g_{\mu\nu} -  \frac {\pa_\mu\pa_\nu }{\pa^2}) \ 
( \slashed \pa -m) \ 
 \hat  \Psi_\nu  +
  (\bar  a {^*} +....) \slashed \Psi .\nn
  $
}
  \bea\label{gfrs}
  L_{RS}^{free} +\bar a ^*\slashed\Psi
  =
    \Psi ^*    _\mu  
  (g_{\mu\nu} -  \frac {\pa_\mu\pa_\nu }{\pa^2})  
 \slashed \pa   
   \Psi_\nu  +
  (\bar a {^*} +....) \slashed \Psi. 
   \eea
 The non-locality  seemingly presents  in  
             $ \pa_\mu\pa_\nu  \frac {{\slashed \pa} } 
   {\pa^2 }$ is   spurious as it will be shown shortly.   Eq.~(\ref{gfrs}) expresses  the naturalness of the off-shell gauge condition~(\ref{gt}).  
   One can define 
   the following off-shell decomposition of  $ \Psi_\mu$
    \bea\label{dec}
   \Psi_\mu= \hat \Psi_\mu   +\gamma_\mu\hat \Psi \quad {\rm where} 
 \quad 
 \hat \Psi  \equiv  \frac1d \slashed \Psi
   \quad  
   \hat \Psi_\mu\equiv   \Psi_\mu-\frac1d\gamma_\mu   \slashed \Psi.
   \eea
          This notation will be   convenient   when     completing the   Unimodularity gauge condition 
   $\sqrt g=1$   by a  $\gamma$-Traceless     condition for~$\Psi_\mu$
   \subsection{Additional fields   for  imposing the $\gamma$ traceless gauge
  in supergravity }
  We wish to  separately gauge fix  
  in  a   BRST invariant  way 
   both  irreducible  spin $1/2$ and spin  $3/2$~spinors 
   $\hat\Psi   $ and $\hat \Psi_\mu$ in Eq.~(\ref{dec})
    with  the      gauge functions
       $\slashed \Psi  $ and  $\pa^\mu \hat \Psi_\mu$.
  This      choice  of spinorial gauge functions  is quite different   than  the conventional ones in supergravity. The latter   amount to  add to the Rarita--Schwinger Lagrangian   a gauge fixing term   $\slashed \Psi ^* \slashed  \pa \slashed \Psi   $ \cite{PVN} with   additional subtleties in the massive case  \cite{BG}.   However, such a  gauge fixing term  vanishes for $\slashed \Psi =0$.  It     is thus  inconsistent with    the  off-shell $\gamma$-Traceless condition  of $\Psi_\mu$  that   will be used shortly   to build  the Unimodular gauge supergravity.   
  
 We must  advance with caution because 
  a    propagator degeneracy   
  for the local supersymmetry ghosts is feared  if we  impose $\slashed\Psi=  0$,  analogous to that     occurring    for  the reparametrization ghosts in  the    gauge   $\sqrt g=1$.
  
    Call $\chi$ the commuting supersymmetry  Majorana spinor  ghost of supergravity. 
 $\bar \chi$  is the commuting antighost  and   $d=s\bar\chi$  is 
  the      anticommuting spinor auxilary field  (often  known as the Nielsen--Kallosh ghost), which is   generally  used to     possibly    enforce a spin  $1/2$ gauge condition on the gravitino. $d$ is the  analog of  $b ^ \mu$  that  allows   a vector gauge condition on the metric.  
    The    standard   BRST symmetry    of the $N=1,  d=4$ supergravity is \footnote {
    For the sake of notational simplicity,    the supergravity auxiliary field dependence  of the BRST transformations
    is left aside. The way to use them for quantization  is eg in \cite{bb} and \cite{BG}. Such refinements  play no  role in the argumentation of this   paper.}
   \bea\label {sbrst}    sg_{\mu\nu}&=& \Lx      g_{\mu\nu}    + i  \Psi^*_{\{\mu}  \gamma_   {\nu\}} \k     \nn\\
  s\Psi _{\mu }&=& \Lx  \Psi _{\mu }  + D_\mu \k     \nn\\
   s\xi^\mu&=&   \Lx   \xi^\mu + i     \chi {^*}  \gamma^\mu\chi   =
   \xi^\nu   \pa_\nu   \xi^\mu   + i     \chi {^*}  \gamma^\mu\chi \nn\\
   s\chi&=& \Lx   \chi=     \xi^\nu   \pa_\nu \chi
    -  \demi \chi
     \pa_\nu  \xi^\nu  \nn\\
     s\bar \chi&=  & d \nn\\ sd &=  &0.
      \eea
      The novelty  will be        the   use   the fermionic Lagrange multiplier 
  $d$ to       gauge fix the irreducible   component $\hat \Psi_\mu$~of~$\Psi_\mu$, and  not the full $\Psi_\mu$. This makes  the  situation quite   different than   for standard gauge choices    of  \cite{PVN} and~\cite{BG}.

Here  is  the point. The  unimodularity constraint  $\sqrt g=1$ implies  for  consistency   that   the   supersymmetry  variation of  $\sqrt {g}$ vanishes. Thus,    
 Eq.~(\ref{co})      generalizes as
  \bea\label{cons}
  0= s \sqrt {g}  |_{\sqrt g=1}=
  \pa_\mu (\sqrt g \xi^\mu) +  i g^{\mu\nu}   \ \chi^*\gamma_\mu\Psi_\nu=
\pa_\mu \xi^\mu +  i   \chi^* \slashed\Psi.
    \eea
  The supergravity  path integral  measure  must therefore separate between   BRST invariant  
     ``transverse" and ``longitudinal"   off-shell field components  of  all ghosts, where ``longitudinal"  means  the conditions $ \pa_\mu \xi^\mu=0$ and  $ \slashed\Psi=0$ and refers to the decompositions  of 
   $g_{\mu\nu} $ and $\Psi_\mu$ in  $\hat g_{\mu\nu} $, $  \sqrt g$,  $\hat \Psi_\mu$ and $\hat \Psi$ .

     For   the Rarita--Schwinger action, 
  to consider $\hat \Psi_\mu$ and $\hat \Psi$ as  the   independent  classical  components of 
  the Rarita--Schwinger field  to be     possibly   gauge fixed    separately, 
 one   must    correspondingly   complete    the standard     supergravity  BRST  fields appearing   in      (\ref{sbrst})  by   addition of   a  spinorial
  trivial    quartet
  $\l,\a,\bar \l,\ab$.  The    reason  is the same as for  having  introduced  the quartet  $L,\eta,\bar \eta,b$   to  possibly gauge fix separately the unimodular part and the conformal factor of the metric, considered as independent field variables of   the  pure  gravity theory and  possibly enforce the Unimodular gauge for    the metric  alone.
    
  The          BRST  gravity   fields     in (\ref{spg})   get therefore the following  Rarita--Schwinger partners       
\bea \label{brst2}
\begin{matrix}
         &   &  \Psi_{\mu }=(\hPsi_{\mu}, \hat \Psi\equiv \slashed \Psi),   \a ^ {(00)}&       \\
    &   \swarrow    \swarrow&  &     \\
   \k^{\mu(10)},   \l  ^ {(10)}  &    &  &  &      \kb  ^ {(01)},     \lb  ^ {(01)}  \\
   &   &   &  \swarrow  \swarrow    \\ 
  &   &   d ^ {(11)},   {\ab}  ^ {(11)}      
    \\ 
     &   &  &    & &   \\
  1   &     &     0  &   &  -1   &    
\end{matrix}. \eea
The  BRST transformations that      complete  those in Eqs.~(\ref{sbrst}) and  express   $\l,\a, \lb,\ab$ as   a trivial      quartet are
 \bea
   \s a   &=& \l      \quad \quad \quad \quad \quad\quad \quad
    \s \l   =0 \nn\\
      \s \lb       &=& \ab   \quad \quad \quad\quad \quad\quad \quad
    \s \ab  =  0. 
      \eea
 \subsection{   BRST  exact-terms for the  $\gamma$-traceless gauge  in   supergravity }
To   impose the $\gamma$-Traceless  condition   $  \slashed{\Psi}=0$ on  the Rarita-Schwinger field,   one defines
\bea\label{gpsi}
L^{\slashed{\Psi} }_{\rm gf} =
  s \Big (
  \bar\l ^*
 \slashed{\Psi}
 )     \Big)=
{ \bar {a}}   ^* 
 \slashed{\Psi}  +   \bar \l ^*   \slashed{D}  \k +   \bar\l {^*}  (se^\mu_a ) \gamma^a  \Psi_\mu.
\eea
The equation of motion of  $\bar a$ enforces $  \slashed{\Psi}=0 $ for the spin  $3/2$ field, analogously as that  of   $b$ enforces   
  $\sqrt g=1$. 

To impose   the  longitudinal  gauge function  $  \pa\cdot \hat \Psi $ on  the $\gamma$-Traceless spin $3/2$ field  $\hat \Psi_\mu$,
 one defines   
  \bea\label{dfpP}
L^{\pa\cdot \hat \Psi}_
{\rm gf} =  s     (    { {\bar \chi }}{^*} (     \pa\cdot \hat \Psi    +  \betar     \slashed{\pa}  a  +{ \frac {\delta  }{2}  \slashed{\pa}    d })
 )
=  {     \delta  d {^*}   \frac{  \slashed  \pa } {2  }  d}+   d {^*}(     \pa\cdot \hat \Psi      
+\betar \slashed{\pa}   a) 
+
 { {\bar \chi }}{^*}   (   \pa\cdot  D  \k    
+  \betar       \slashed{\pa}    \l)
\nn\\
+ {\rm{ ghost\  interaction\  terms}\ proportional \ to \ }sg_{\mu\nu}.
\eea
   $\betar$  and{     $\delta$ are parameters}.    The    field  $a $ is  the fermionic analogous of the boson $L$ in   Eq~(\ref{Egf}).
   
     The proposed  gauge fixed action of the massless Rarita--Schwinger field in  the $\gamma$-traceless gauge  is  therefore
\bea\label{RSgf}
\int dx L^{\rm RS}_{\rm BRST} ( \Psi_\mu, \chi,\bar\chi, d, \l,\bar\l, a,\bar a)
\equiv
\int  dx (\ 
L^{RS}+L^{ \slashed{\Psi}   }_{\rm gf} +
L^{\pa\cdot \hat \Psi}_
{\rm gf}).
\eea
The    $\gamma$-traceless  condition  (\ref{gt})  $\slashed \Psi =0$   holds everywhere after     the elimination of   
$\bar a$  by its algebraic equation of motion from  $L^{\rm RS}_{\rm BRST} $, while the corresponding  BRST symmetric ghost term remains. In particular, the   
  free quadratic part of  $ \int dx  L^{RS}$   is   gauge fixed to
  (\ref{gfrs}).
  One must check that      all fields  in  (\ref{brst2})  have invertible propagators stemming from the action (\ref{RSgf}). 

The    term $ s ({ {\bar \chi }}{^*}    \slashed{\pa}  a)$   in  $L^{\pa\cdot \hat \Psi}_
{\rm gf}$  enforces  the  propagation of the   fields  $\l$ and $a$.  In order its coefficient doesn't vanish, one has the following  condition, analogous to $\gamma\neq0$ in (\ref{Egf}),
\bq \betar\neq  0.\eq

  $L^{\pa\cdot \hat \Psi}_
{\rm gf}$    determines    a mixed   propagator between  $\pa^\mu \hat \Psi_\mu$ and  the  spin $1/2$ 
 field  $d$.
The    second order    propagation term ${^*} \kb   \pa^2  \k$   between   the   supersymmetry  ghosts 
$ \k  $  and 
$ \kb   $   
 is a  mere  consequence of the  choice   of a  gauge fixing  function  $\pa\cdot \hat \Psi$ to gauge fix the remaining of 
 the   local supersymmetry invariance after having imposed $\slashed \Psi=0$.

 
\subsection{Free quadratic approximation of  
 the  $\gamma$-traceless gauge  fixed   BRST invariant  Rarita--Schwinger action }
{\blue The equations of motion  of $ \int dx L^{\rm RS}_{\rm BRST} ( \Psi_\mu, \chi,\bar\chi, d, \l,\bar\l, a,\bar a)$  present    a complicated aspect analogous to that    already discussed for the action  (\ref{LBRSTg}) of   the Einstein theory  in the Unimodular gauge. The analysis of each term in $ \int dx L^{\rm RS}_{\rm BRST} $ can be done   as we did   in section   2.2  for    the  action  (\ref{LBRSTg}).  To verify that the gauge fixing is complete and consistent,  it is enough to display    the quadratic approximation of the   Lagrangian, out of which one can get a clear  insight  on the general aspect  of the  equations of motion. One has }
\bea
\int  dx \Big( 
L^{RS}_{\rm free} 
+L^{ \slashed{\Psi}   }_{\rm gf} +
L^{\pa\cdot \hat \Psi}_
{\rm gf})
\Big )
\equiv  \int  dx \Big( 
 L_{\rm free} ^F   (\Psi_\mu, a, \bar a) +    L_{\rm free} ^B(  \chi, \bar \chi,  \l,\bar\l ) \Big ).
\eea 
 The fermionic part of $L_{\rm free} $ is  
  \bea\label{matrixf} 
  L_{\rm free} ^F   &=&
 ( \bar a {^*} +...) \slashed{\Psi}  +
  \eta^{\mu\nu}           \Psi ^*    _\mu  \slashed \pa  \Psi_\nu 
  -
  \pa\cdot     \Psi ^*      
    \frac {1}{ \slashed \pa } 
   \pa \cdot   \Psi{^*}
  +\delta d {^*}  \frac{  \slashed  \pa }   {2 }  d
  +  d {^*}     ( \pa\cdot  \Psi    +  \betar     \slashed{\pa} a )   
\nn\\
&\sim& 
  \eta^{\mu\nu}          \hat\Psi ^*  _\mu  \slashed \pa \hat \Psi_\nu 
   -
  \pa\cdot     \hat\Psi ^*    
    \frac {1}{ \slashed \pa } 
   \pa \cdot      {\hat {\Psi}} 
  +\delta d {^*} \frac{ \slashed \pa } {2} d 
  +  d {^*}     ( \pa\cdot  \hat \Psi    +  \betar     \slashed{\pa} a )  
\nn\\
&\sim& 
  \eta^{\mu\nu}          \hat\Psi ^*  _\mu  \slashed \pa \hat \Psi_\nu  
 -
  \pa\cdot     \Psi ^*      
    \frac {1}{ \slashed \pa } 
   \pa \cdot   \Psi
 -   ( \pa\cdot  \hat \Psi^*    +  \betar     \slashed{\pa}a^*  )    \frac{ \delta } {\slashed \pa}   ( \pa\cdot  \hat \Psi    +  \betar     \slashed{\pa} a )  .
\eea
In both  last  lines,  the    Rarita--Schwinger field dependance  is only through its    spin  $3/2$ $\gamma$-traceless  component~$\hat   \Psi_\mu$~after    eliminating    $\bar a$ by its equation of motion.   
Taking $\delta = -1$, the terms  $ \pa\cdot   \hat  \Psi^*   
    \frac {1}{ \slashed \pa } 
   \pa \cdot     \hat\Psi   $ cancel and  one gets the following     Lagrangian that defines     the   fermionic   free propagators  of    $\hat   \Psi_\mu$ and  $a$ with      a mixing   for $\betar\neq0$
    \bea
  L_{free,\delta = -1 }^F\sim
  \eta^{\mu\nu}          \hat\Psi ^*  _\mu  \slashed \pa  \hat \Psi_\nu  
+\betar^2    a^*     \slashed{\pa} a 
+2\beta     a  ^*   \pa\cdot\hat \Psi.
 \eea
 The bosonic  part of $L_{\rm free} $ is
 \bea\label{bapro}
  L_{free,\delta = -1 }^B=
\begin {pmatrix} 
 \bar \chi {^*}     &   \bar\l {^*}  \ 
\end {pmatrix}
\begin {pmatrix}\ 
\pa^2 
     &\ \betar   \slashed{\pa}    \cr  
  \slashed{\pa}   & 0 \end {pmatrix}\begin {pmatrix}\ 
\chi    \cr  \l  
\end {pmatrix}.
\eea
The choice  $\betar\neq0$  is necessary for the invertibility, giving   the following   matrix of   free bosonic propagators 
    \bea\label{baproi}
\begin {pmatrix} 
  \bar \chi {^*}      &    \ab {^*}    
\end {pmatrix}
\begin {pmatrix}\ 
0
    &  \frac{1}{ \slashed{\pa} }   \cr  
\frac{1}{ \betar  \slashed{\pa}  }   & - \frac {1}{\betar }    \end {pmatrix}\begin {pmatrix} 
\chi   \cr  \l  
\end {pmatrix}.
\eea
The  propagators stemming from $L_{\rm free} $ have    standard  dimensions and are suitable for a perturbative expansion. 
The constraint  $\slashed \Psi=0$  holds  in the Feynman rules of interactions. 
The  spin $1/2$ component  $\slashed \Psi$ of the Rarita--Schwinger field  doesn't circulate   within   loops. This phenomenon      is compensated by a circulation of   appropriate ghosts. The   decoupling of   $\slashed \Psi$  and    of  the conformal factor   $\phi$ are  analogous   phenomena in the Unimodular gauge.

%
%
   
\subsection{Supergravity action in the Unimodular gauge }
In  the Unimodular   gauge for the graviton and   $\gamma$-Traceless  gauge for     the gravitino,   
the previous results give   the following   
   BRST invariant   gauge fixed  action for  
  the classical supergravity action    $ \int dx
\sqrt {g } (R( g_{\mu\nu}    ) 
+\demi  i
  \epsilon^{\mu\nu\rho\sigma}\ 
   \Psi_\mu {^*}  \gamma^5  \gamma_\nu  D_\rho \Psi_\sigma  )
   $
\bea 
I_{\rm supergravity } ^{\rm Unimodular}  [   \hat g_{\mu\nu} , \hat \Psi_\mu, {\rm ghosts }]=
  \int dx \Big (
  R( \hat g_{\mu\nu}    ) 
+\demi  i
  \epsilon^{\mu\nu\rho\sigma}\ 
      \hat\Psi ^*  _\mu  \gamma^5  \gamma_\nu  D_\rho \hat \Psi_\sigma \Big )  +{\rm ghost \ terms}.\eea 
The fields $b$ and  $\lb$   have been eliminated    by  their algebraic  equations of motion.  This    BRST invariant action depends  on the metric and   on the  Rarita--Schwinger  field     only  through   their  unimodular and   $\gamma$-Traceless   components  $\hat g_{\mu\nu} $ and  $ \hat   \Psi _\mu$. 
     Such a  genuine    dependence  in function of the metric and  the  gravitino              simplify     the  expression    of  Ward identities of local supersymmetry.
  In particular,   all  terms obtained by  variations of~$\hat g_{\mu\nu}$    are   traceless  and     interactions  between     spin $1/2$ and  $3/2$ components  of 
     $  \Psi_\mu$  disappear because~$\slashed\Psi=0$.
     
     { \blue
    
     \subsection{  The      first order   formalism  spin connection in the Unimodular gauge  }

   Explaining  the determination  of  the spin connection $\omega$     in the     Unimodular   gauge  is necessary   to   make   precise how all half-integer spin fields couple   in this gauge   to    gravity and supergravity  via  their covariant derivative  $D=d+\omega$. 
      In the     first order formulation,    the  1-form vielbein $e^a= e^a_\mu(x)$  and      the 1-form spin connection $\omega^{ab}(x) =\omega^{ab}_\mu (x)  dx^\mu$ that gauges the local Lorentz transformations
      are  introduced as  independent fields  ($a,b...$ are  Lorentz indices).    
      For any given  supergravity model expressed in first order formalism, one   generally  
   eliminates  the  spin connection by its  algebraic equation of motion   and expresses it   as a local function of the metric, the gravitino and possibly       auxiliary fields.
   Equivalently, one can impose  a  covariant constraint on the torsion  in the supergravity  first  order action,   solved by the same choice of the spin connection \cite{PVN}.  In fact the determination of the spin connection  by   such a  covariant constraint respects the off-shell closure of the gauge symmetries provided  the constraint on the torsion is  fixed    such   that it is compatible with   Bianchi identities \cite{bb}.

    In what follows, we keep restricting to  the      $N=1,d=4$ supergravity case and     neglect  for simplicity  the auxiliary fields  (see however the footnote $^{**}$ for their inclusion).
   In this case,  the  curvature  $R^{ab}$ of the Lorentz gauge field $\omega_{ab} $, the  torsion   $T^a$  of $e^a$  and the field strength $\rho$ of the   
 gravitino 1-form  $\Psi =\Psi_\mu dx^\mu $ are the following  2-forms satisfying Bianchi identities that amount to $d^2=0$ when all curvatures vanish
\bea
   R^{ab} 
   &\equiv&
    d\omega^{ab}+\omega^a_c \omega^{cb}
   \nn
  \\
   T^ a    &\equiv&  de^a+\omega^a_b e^b +\frac {i}{4}   \Psi^* \gamma^a \Psi\nn
  \\
  \rho &\equiv& D\rho =d\Psi + \omega \Psi = d\Psi + \omega ^{ab}  \sigma_{ab} \Psi.
   \eea
    ($\sigma_{ab}\equiv \frac i 2 [\gamma_a,\gamma_b])$. The first order supergravity action is
    \bea\label{sugraIlIl}
    I_{\rm first\ order}=
   \int ( \epsilon_{abcd } \      e^a\wedge e^b  \wedge R^{cd} 
   +
   \frac i 2  \Psi ^* \gamma ^5  \gamma_a  \wedge e^a \wedge \rho)
  .
   \eea
   The   equation of motion of the spin connection $\omega$ of this    action  is the super-Poincar\'e   torsion zero condition
   $  T^a=de^a+\omega^a_b e^b +\frac {i}{4}    \Psi\gamma^a \Psi=0
   $.  
   As said earlier, this condition can   be  postulated   from the beginning as a geometrical constraint (with a possible     covariant distortion   when      auxiliary fields are introduced compatible with the invariances of 
   the action ({\ref{sugraIlIl})\footnote{For including the auxiliary field dependence in the discussion, \cite{bb} shows the way to go.  When auxiliary fields  are introduced to get a closed system of equations without using some equations of motion, the classical supergravity action becomes
 $ \int ( \epsilon_{abcd } \  \  e^a\wedge e^b  \wedge R^{cd} +
   \frac i 2  \Psi ^* \gamma ^5  \gamma_a  \wedge e^a \wedge D_\omega\Psi 
   -B_2\wedge dA +      G_3 \wedge \tilde G_3
   )$  where     $G_3=dB_2 +\frac i 4 \Psi ^* \gamma_a \Psi e^a$     is the curvature of the auxiliary 2-form. The gauge symmetries   are   distorted but  an algebraic    equation of equation  on $\omega$ still holds  through the covariant  torsion condition  $T^a =-\demi G_{abc} e^be^c$. This  change  only  modifies   the relation between the spin connection $\omega _{abc}$  and the vielbein and gravitino by  terms proportional to the    components   $G_{abc}$ of the 3-form curvature.  In the Unimodular gauge one   gets 
  $ G_3\to   \hat G_3=dB_2 +\frac i 2  \hat \Psi ^* \gamma_a  \hat \Psi \hat e^a$ and Eq.~(\ref{spincc})   
 $\to  \omega^{ab}_\mu = \hat   \omega_\mu^{ab}\equiv 
   \omega_\mu^{ab} (\hat e^a_\mu,\hat g_{\mu\nu},  \hat \Gamma_{ \mu\nu\rho} (\hat g_{\alpha\beta}), \hat \Psi_\mu, A_\mu, G_{abc})$.}). 
   The constraint  on   $T^a$   
         is an  invertible   system   of linear equations  for  the components 
    $\omega^{ab}_\mu$.     The $d(d-1)/2$ dimensional local Lorentz symmetry   of    $I_{\rm first\ order}$     can  be gauge   fixed  by  imposing the  $d(d-1)/2$ relations 
$e^a_\mu=e^\mu_a$, giving a one to one one correspondence between the $d(d+1)/2$  components of the Lorentz  gauge-fixed $e^a_\mu$ and  those of $g_{\mu\nu}$, by   using    $g_{\mu\nu} \equiv e^a_\mu e_{a\mu}$. The  Lorentz  gauge-fixing  gives a trivial Fadeev--Popov determinant implying    the consequent-less  elimination of the  Faddeev--Popov ghosts of the local Lorentz symmetry  by their algebraic    equations of motion stemming from  by the BRST symmetry.  Their solution is      a (complicated)   local   function  of   the 
 $d(d+1)/2$ non vanishing components of    $e^a_\mu$, of  their   derivatives and  of the gravitino    $\Psi_\mu$. In fact, as shown with many details   for instance in \cite{PVN},  one has   explicitly     
 \bea\label{sc}
T^a=de^a+\omega^a_b e^b +\frac {i}{4}    \Psi\gamma^a \Psi=0 \quad \rightarrow  \quad
 \omega_\mu^{ab}   =\omega_\mu^{ab} (e^a_\mu,g_{\mu\nu}, \Gamma_{\mu\nu\rho} (g_{\alpha\beta}),  \Psi_\mu).
 \eea 
The basic property   of  the first order  formalism is  thus    the equivalence of both following  local   actions, where the gauge field  $\omega$ of the Lorentz symmetry  in  the right  hand side by is     computed in    Eq.(\ref{sc}),    
    \bea\label{sugraIl}
   \int \  \epsilon_{abcd } \   e^a\wedge e^b  \wedge R^{cd} (\omega) +
   \frac i 2  \Psi ^* \gamma ^5  \gamma_a  \wedge e^a \wedge D^{(\omega)}\Psi 
   \sim
    \int dx \sqrt g\ 
    ( \ R^{(\omega(e,\Psi))} + \frac i 2\epsilon^{\mu\nu\rho\sigma}
   \Psi ^*    _\mu  \gamma^5  \gamma_\nu  D^{(\omega(e,\Psi))}_\rho \Psi_\sigma \ ).
\eea
The Unimodular gauge  that fixes    $g_{\mu\nu} \to  \hat g_{\mu\nu} $  and $\Psi_\mu \to \hat \Psi_\mu$  is  obtained by   further    imposing   $\phi=0$ and $\hat \Psi=0$ everywhere,    adding to the action  the   BRST exact terms
discussed   above.  Notice that, in  first order formalism,    the gauge condition $\phi=0$~implies~$e^a_\mu  \to  \hat e^a_\mu,  $ where $ \hat e^a_\mu  $  can be  parametrized by $d(d+1)/2-1$ fields    because of the relation 
$\hat g_{\mu\nu} = \hat e^a_\mu \hat e_{a\nu}$ and $\det \hat g_{\mu\nu} =1$.  Moreover,  with $\phi=0$ the   vanishing torsion condition implies   $  d\hat e^a+\omega^a_b\hat  e^b +\frac {i}{2}   \hat  \Psi^*\gamma^a \hat \Psi=0
   $. The   solution for $\omega$ is     as in  (\ref{sc})  by     replacing   $g_{\mu\nu}$,  $e^a_\mu$, $\Psi_\mu$ by  $\hat g_{\mu\nu}$, $\hat e^a_\mu$, $\hat\Psi_\mu$,  so   that  
 \bea \label{spincc}   \omega^{ab}_\mu = \hat   \omega_\mu^{ab}\equiv 
   \omega_\mu^{ab} (\hat e^a_\mu,\hat g_{\mu\nu},  \hat \Gamma_{ \mu\nu\rho} (\hat g_{\alpha\beta}), \hat \Psi_\mu).
\eea
  This formula     of the spin  connection   in  the Unimodular gauge   generalizes  the range of applications of the golden rule of \cite{bcw}  to the first order formalism.   Using Eq.~(\ref{spincc}), the expression of the covariant derivatives $d+\omega$ is what  determines the details  all  half-integer spin couplings in the Unimodular gauge. 
    }}

\section{Conclusion}

 The  completion  of the ordinary BRST field content of supergravity  
$  (g_{\mu\nu},\xi^\mu,\bar  \xi^\mu, b^\mu)$ and  
$  (\Psi _\mu,\chi, \bar\chi,d ) $
by the pair  of   both  BRST trivial quartets counting  for zero degrees of freedom 
 \bea ( L,\eta,\bar \eta,b) \quad \rm {and} \quad (\l,\a,\bar \l,\ab)\eea   
 allows  an  off-shell      BRST invariant  gauge fixing  of   the  metric   and  Rarita--Schwinger fields      into their unimodular part  $\hat g_{\mu\nu}$ 
   and  $\gamma$-Traceless part  $ \hat \Psi_\mu$.  
   
    This      Unimodular gauge choice     
   $\sqrt g=1 $ for the metric and   the~$\gamma$-Traceless  condition  $\slashed\Psi=0$  for the Rarita--Schwinger   field
  can be   further  completed by the  less qualitative gauge functions
   $\pa^\nu\hat g_{\mu\nu} $ and $\pa^\mu\hat \Psi_\mu$.   One gets an off-shell decoupling of the conformal factor of the metric and    of the  $\gamma$-Trace of the    Rarita--Schwinger field.
   This new class  of gauges   gives  a different and maybe   quite   interesting  perturbative theory of  gravity and supergravity where the conformal factor $\phi$ is gauge fixed to zero ab initio in a BRST invariant way.{\blue  The paper also indicates   the relation satisfied  by   the   first  order   order formulation  spin connection  in the Unimodular gauge.}

  A   virtue of our extended BRST analysis is to provide a clearer approach to the definition of  observables in gravity and supergravity. 
   In view of Eq.~(\ref{observable}) (and its   extension to supergravity), observables  can    be defined as functionals 
of $\hat g_{\mu\nu}$
   and   $\hat \Psi_\mu$ in the cohomology of $s$.    The   set of physical  $S$-matrix elements   to   be  computed in this gauge are  
     those     with  external  legs made of    unimodular components of the graviton  and $\gamma$-traceless gravitino.
   Getting  the conformal factor   and the  spin~$1/2$  component of the gravitino  as  spectators 
      extends at the quantum level the  old   classical intuition  of Einstein~\cite{Einstein} and  the     work of  \cite{York}  at least semi-perturbatively.

   \cite{bb}      observed      that a  $2$d  world sheet is  best described in terms   of the  Beltrami parametrisation of   the  $2$d metric (and  the $2$d gravitino), with    a  correspondence between  the 2d  unimodular metric  $\hat g_{\mu\nu}$ and     the Beltrami differential $ \mu^z_{\bar z}$.
Interestingly,  the present paper generalizes   to all dimensions  $d>2$
 the possibility of formulating gravity (and supergravity) with   the reduced  fields $\hat g_{\mu\nu}$  (and $\hat \Psi _\mu $) as fundamental fields, modulo some ghosts that are  not in the physical spectrum. 
  In the  $2$d  case, this gives a precise   understanding of     the    factorisation properties of (super)strings,   the  decoupling  of the conformal factor  of  the  world sheet,       
 the nature  of    $2$d~(super)conformal  anomalies,  the definition  of (super)string observables and so on. The   $2$d~(super)conformal factor fully  disappears from the path integral measure and the Liouville fields      couple only  to  the (super) Beltrami components of the   $2$d metric (and  $2$d gravitino). 
  The perspectives of using the Unimodular  gauge for $d>2$  are not yet obvious.   It might  be for instance illuminating to revisit  the Velo--Zwanziger phenomenon~\cite{vz} as well as  the BRST   superHiggs effect analysis  of  \cite{BG}  in this different  gauge for the Rarita--Schwinger field.  
       In fact, given  that any  perturbation around an unimodular background is purely traceless, as in particular   a classical  graviton is,  one may consider   the Unimodular gauge  as a kind of physical gauge for gravity. Reformulating known General Relativity solutions  in this gauge might  be quite instructive.     
%
 \bigskip

\bigskip\noindent {\bf Acknowledgments:}
It  is a pleasure  to thank Jean-Pierre  Derendinger,  John Iliopoulos and Mathias Blau for      interesting   discussions on the subject of this paper.

\end{document}